\documentclass[12pt]{article}

\date{}

\begin{document}
\thispagestyle{empty}
\title{\bf An explanation why the $\Theta^+$ is seen \\
in some experiments and not in others}
\author{Ya. Azimov$^{1}$, K. Goeke$^2$, I. Strakovsky$^3$\\ \\
$^1$ Petersburg Nuclear Physics Institute,\\ Gatchina, 188300 Russia \\
$^2$ Institut f\"ur Theor. Physik -II, Ruhr-Universit\"at,\\
D-44780 Bochum, Germany\\
$^3$ Center for Nuclear Studies, Physics Department, \\ The George
Washington University,\\ Washington, DC 23606 USA} \maketitle

\begin{abstract}
To understand the whole set of positive and null data on the
$\Theta^+(1530)$-production, we suggest the hypothesis that
multiquark hadrons are mainly generated from many-quark states,
which emerge either as short-term hadron fluctuations, or as
hadron remnants in hard processes. This approach allows us to
describe both non-observation of the $\Theta^+$ in current null
experiments and peculiar features of its production in positive
experiments. Further, we are able to propose new experiments that
might be decisive for the problem of the $\Theta^+$ existence.
Studies of properties and distributions of the $\Theta^+$ in such
experiments can give important information on the structure of
both conventional and multiquark hadrons. It would provide better
insight into how QCD works.


\end{abstract}

\section{Introduction}

Invented more than 40 years ago~\cite{GN,GM,Zw}, quarks were initially
introduced rather formally, to account for a very limited variety
of hadronic flavor multiplets known at that time. Their simplest
application was to present every baryon as a three-quark system
and every meson as a quark-antiquark system. Then, in respect to the
flavor symmetry group $SU(3)_F$, the mesons could exist only as
singlets and/or octets, while the baryons could reveal both the same
kinds of multiplets and, in addition, decuplets.

Such a simple conventional picture was in good correspondence with
experiment.  It could be quite successful, if the quarks were nothing
more than mathematical objects allowing to visualize symmetry
classification of hadrons (such treatment would not contradict to
Gell-Mann's paper~\cite{GM}). However, if quarks (or ``aces'', in
Zweig's terminology~\cite{Zw}) are physical objects, the picture of
hadrons with a fixed number of constituents could be self-consistent
only in a non-relativistic theory. In a relativistic case, production
and annihilation of virtual quark-antiquark pairs prevent the total
number of quarks and antiquarks from being fixed.  That is why the
first hints, that the quarks might be something more than just
mathematical objects, revived the question whether one would be able
to observe hadrons with non-conventional quark content, entering some
other kinds of flavor multiplets and, in particular, having exotic
quantum numbers, impossible in the conventional picture.

One of the simplest clear examples of non-3-quark baryons would be
a baryon with positive strangeness, say, a resonance in the $KN$
system with $S=+1$ (by definition, the strange quark $s$ has
$S=-1$, and a 3-quark system cannot have $S=+1$). Experimental
searches for such exotic states in the $KN$-spectra started rather
early~\cite{cool,abr,tys}. However, all suggested evidences stayed
unconvincing, and later the Particle Data Group (PDG) stopped
discussions on experimental spectroscopy of exotics~\cite{PDG86}.

The first theoretical attempt to describe internal dynamical
structure for specific multiquark hadrons was made in the
framework of the ``MIT bag"~\cite{jaf1,jaf2,jaf3}. The
calculations supported existence of exotic states, but prescribed
them to be very broad, with widths of some hundreds MeV. This
result seems quite understandable, since an exotic multiquark
system in the MIT bag looks to be readily prepared for separation
into subsystems with conventional quark contents: a tetraquark
meson $qq\overline{q}\overline{q}$ may be considered as a system
of two quark-antiquark pairs, while the pentaquark baryon
$qqqq\overline{q}$ may be considered as (3$q$ + $q\overline{q}$).
According to the MIT-bag approach, it is just the enormously huge
width that could explain why exotic resonances have not been seen in
experiment: they exist, but are too broad to reveal clear-cut bumps.
Totally unnoticed for long time has stayed an alternative possibility:
the fact that hadrons have an internal structure may suggest some
hadrons to have a very complicated structure which would suppress their
couplings (and, therefore, production cross sections and decay widths)
to conventional hadrons~\cite{az70}. It could be similar to
``structural'' suppression of some radiative transitions, which is
well-known in atomic physics. Such possibility has been recently
demonstrated by detailed calculations for the system of
$(4q)\overline{q}$~\cite{cc}.

It is interesting, that the standard partial-wave analysis of the
elastic $KN$-scattering~\cite{hys} (the latest and most complete
one published up to now) has later presented four exotic baryon
states (resonances), two isoscalar and two isovector, all they
having, indeed, large widths 200--500~MeV. However, even before
the publication~\cite{hys}, any correspondence between the MIT
bag and experiment had become to look generally dubious.

New impetus for studies of exotics has emerged~\cite{dpp} from
the Chiral Quark Soliton Approach ($\chi$QSA; for recent reviews
see Refs.\cite{wk,ekp}). It allowed to give rather detailed
theoretical prediction of the exotic antidecuplet of baryons
(including the $\Theta^+$ with $S=+1$) and strongly stimulated
new experimental searches for exotics. These efforts gave, at
last, some positive evidences, as summarized in the Review of
Particle Properties, issue of 2004~\cite{PDG04}.

Nevertheless, the current experimental status of the exotic
baryon $\Theta^+(1530)$ is rather uncertain (see the more recent
experimental review~\cite{bur} and the latest edition of the
Review of Particle Properties~\cite{PDG06}). Some collaborations
give positive evidences for its observation, some others do not
see it, thus casting doubts on existence of the $\Theta^+$ and on
correctness of its positive evidences. Especially impressive are
the recent high-statistics null results of the CLAS Collaboration
for $\Theta^+$-photoproduction into several final-states~\cite{clas1,
clas2,clas3,clas4}. And yet, new (``after-CLAS'') dedicated analyses
have again provided both confirming~\cite{leps,svd,svd-m,bar,tr} and
null~\cite{foc, l3, ank,tof,nom} publications. Moreover, a new
suggested way~\cite{adp} for data analysis may reveal presence of
the $\Theta^+$ even in the published CLAS results, through its
interference with a known resonance.

At present, an important fact is that there are no data sets,
from independent groups, with exactly overlapping observational
conditions (initial and/or final states, kinematical regions).
Therefore, when comparing today different current data on the
$\Theta^+$, one needs some theoretical models/assumptions for
the unknown production mechanisms.  Thus, one cannot yet reject
experimental existence of the $\Theta^+(1530)$ as a particular
exotic pentaquark baryon. On the other side, we emphasize also
that, up to now, there has not been suggested, in the framework
of Quantum Chromodynamics (QCD) or in some other terms, any
theoretical reason to forbid multiquark hadrons.

In such uncertain situation, Karliner and Lipkin~\cite{kl,lip}
raised the question shown in the title here, why the $\Theta^+$ is
seen in some experiments and not in others. As an answer, they
suggested an important role of the baryon resonance with hidden
strangeness $N(2400)$ (see also Ref.\cite{as}). However, this
particular resonance production seems to be insufficient today
to explain many of positive evidences.

An opposite viewpoint is that all positive results might arise
as statistical fluctuations and do not reveal a true physical
effect (see, \textit{e.g.}, Ref.\cite{hic}). However, it would
be strange to have the same fluctuation in data of more than
ten independent groups studying very different processes. Moreover,
in such a case we should live with the open question of what prevents
exotic hadrons from being existent.

Therefore, we assume in the present paper that the $\Theta^+$,
as a representative of exotic hadrons, does exist and has
properties corresponding to the published positive evidences:
rather low mass and unexpectedly narrow width. Then we are going
to reconsider possible dynamical picture for multiquark hadrons.
Though our approach is still qualitative, not quantitative yet, it
seems to reconcile different data amd explain some regularities in
(non)observation of the $\Theta^+$. It allows, further, to suggest
new experiments for confirmation and more detailed studies of the
$\Theta^+$-production and its properties.

\section{Experimental $\Theta^+$-production}

Up to now, positive evidences have been published for three exotic
baryons: $\Theta^+(1530),~ \Theta_c^0(3100)$ and
$\Xi_{3/2}^{--}(1860)$ (or $\Phi^{--}(1860)\,$)~\cite{PDG06}. Each
of the two latter states was seen by one group only; they have not
been found in other dedicated experiments. That is why we will not
discuss them here. More crucial in the present situation looks the
existence or non-existence of the $\Theta^+$. The corresponding
information (positive or negative) is much more copious than for any
other exotic hadron candidate.

When considering the $\Theta^+$ to be a real object, we encounter
a problem, whether all existing positive and null data may be
mutually consistent. This seems doubtful under assumption of familiar
hadroproduction mechanisms. However, a very small width of the
$\Theta^+$ provides a hint of an unfamiliar mechanism for the
$\Theta^+$-decay. If so, the production mechanism, most probably,
should be unusual as well.

Up to now, only some of data, both published and preliminary, seem
to touch not only a problem of the $\Theta^+$-existence, but also
a possible mechanism of its production. It is natural, therefore,
to pay special attention just to such data. We will discuss the
positive and negative experimental data separately.

\subsection{Positive data}

a) \textit{ZEUS data.} Let us begin with results from the ZEUS
detector at HERA on the $\Theta^+$-production in the deep-inelastic
scattering (DIS). They are most advanced in respect to the
production mechanism.

As is known, DIS is a typical hard process. At the parton level,
it essentially corresponds to the virtual photon knocking-out a
parton from the initial hadron (proton, in the case of HERA).
Respectively, the final hadrons in DIS may be separated into two
parts: the current fragmentation region, with final hadrons mainly
produced in hadronization of the knocked-out parton, and the
target fragmentation region, where final hadrons come mainly
from hadronization of the target remnant. Note that the latter
case has never been described from first principles of QCD (see,
however, some attempts~\cite{gdkt}).

The ZEUS Collaboration has found both the $\Theta^+(1530)$
and its antiparticle~\cite{zeus1}~\footnote{Note that the
H1 Collaboration, also at HERA, does nor see the
$\Theta^+(1530)$~\cite{h1}. However, both ZEUS and H1 Collaborations
agree that their data do not contradict each other. Recent analysis
with comparison of the two data sets is given in Ref.\cite{ChL}.}.
They are seen in the kinematical region, which was believed to be
related with fragmentation of the knocked-out parton(s)~\cite{exp,kar}.
The proton remnant has been usually considered to escape nearly
unnoticed by the ZEUS detector. In any case, the remnant contribution
in the region used was assumed to be negligible. Such expectations have
been confirmed by measurements of fragmentation functions for various
identified hadrons. In particular, the ZEUS Collaboration measured
fragmentation fractions for different charmed hadrons in this
kinematical region~\cite{cop}. They coincide quite well with those
measured in $e^+e^-$ annihilation. Such coincidence was predicted for
DIS in the current fragmentation region (fragmentation of the
knocked-out (anti)quark is expected to be nearly the same as
fragmentation of the (anti)quark produced in $e^+e^-$
annihilation)~\cite{gdkt}.

Special attention was given to production properties of the baryons
$\Lambda(1520)$ and $\Lambda_c^+(2285)$, since they could be
kinematically similar to production of the $\Theta^+(1530)$. It
appears~\cite{zeus2} that the $\Lambda(1520)$ and
$\Lambda_c^+(2285)$ are produced as expected: $\Lambda(1520)$
through fragmentation of the knocked-out quark, while
$\Lambda_c^+(2285)$ through fragmentation of the
$c\overline{c}$-pair, generated by the $\gamma^*$-gluon fusion.

However, $\Theta^+(1530)$ in the same kinematical region clearly 
demonstrates distributions which are characteristic for hadronization 
of the target-proton remnant (in particular, it is mainly produced in 
the forward hemisphere, \textit{i.e.}, at positive pseudorapidity 
$\eta$)~\cite{zeus2}. Thus, even in the region, that seems 
kinematically related to the current fragmentation, contribution of 
the knocked-out parton(s) to the $\Theta^+(1530)$-production apears 
to be small, if present at all. This is a very essential difference 
between processes of producing the exotic $\Theta^+(1530)$-baryon or 
various conventional 3-quark baryons, such as $\Lambda(1520)$ and 
$\Lambda_c^+(2285)$.

The proton remnant in DIS is always a mixture of various many-quark
configurations, contrary to a few-quark system of the knocked-out
parton(s). Thus, the above ZEUS data allow us to suggest the hypothesis
that the $\Theta^+$-production comes mainly from hadronization of
many-quark (or, more generally, many-parton) systems.

With such hypothesis, we should expect that the $\Theta^+$-production
in DIS may change with $Q^2$, due to changing role of different
many-parton configurations.  Experimentally, the absolute
$\Theta^+$-yield in the ZEUS data decreases with increasing $Q^2_{\rm
min}$ at $Q^2>20$~GeV$^2$~\cite{kar}. However, it is mainly due to the
factor $1/Q^4$ coming from the photon propagator squared. This
universal factor has no relation to remnant configurations and can be
eliminated, if one considers the relative yield of the $\Theta^+$ in
respect to conventional hadrons. The ZEUS Collaboration investigated
the ratio $\Theta^+(1530)/\Lambda(1116)$ in the same kinematical
region. Instead of decrease, this ratio shows slow increase, though,
with current large uncertainties, it may be considered also as a
constant, about 4\%~\cite{kar}.

Such situation reminds the case of nucleon structure functions, which
initially seemed to be $Q^2$-independent (Bjorken scaling). But later,
both theoretical considerations and more precise measurements have
revealed scaling violations in the structure functions, with slow
(logarithmical) dependence on $Q^2$. Since the underlying physics, the
enhanced role of many-parton configurations at higher $Q^2$, is the
same in these two cases, we expect that the ratio
$\Theta^+(1530)/\Lambda(1116)$ should also logarithmically change
(increase?) at high $Q^2$. Evidently, it is a prediction for future
experiments, which can, thus, check our hypothesis.

Meanwhile, we would like to discuss one more feature of the ZEUS
data. The $\Theta^+$-peak is not seen in the $p K_S $ spectrum at
$Q^2>1$~GeV$^2$~\cite{zeus1}. In the HERA kinematics, one has
\mbox{$W^2<10^5$~GeV$^2$,} where $W$ is the $(\gamma^*p)$ c.m.-energy,
and the above restriction for $Q^2$ may be rewritten for the Bjorken
variable $x=Q^2/(W^2+Q^2)$ as $x>10^{-5}$. A very interesting point is
that the ZEUS data do demonstrate the $\Theta^+$-peak even at
$Q^2>1$~GeV$^2$, if one applies additional restriction
\mbox{$W<125$~GeV~\cite{zeus1}, \textit{i.e.},} \begin{equation}
\label{x_B} x > 6.4\cdot10^{-5}\,.  \end{equation} It is well known, on
the other side, that nucleon structure functions increase at very small
$x$. Therefore, the larger contributions to the DIS inclusive
hadroproduction (in particular, to the $p K_S$ continuum spectrum) at
$Q^2>1$~GeV$^2$ should come from the lower region \begin{equation}
\label{x_Bl} 10^{-5} < x < 6.4\cdot10^{-5}\,, \end{equation} which
violates condition (\ref{x_B}).  Thus, the ZEUS data~\cite{zeus1} imply
that the $\Theta^+$-production in DIS depends on $x$ and in the region
(\ref{x_Bl}) should not have such enhancement (if any), which appears
at small $x$ for the usual DIS production of the background
non-resonant $p K_S$ system. Note that, kinematically, the region
(\ref{x_Bl}) cannot be reached at HERA in other $Q^2$-regions
investigated by ZEUS (more exactly, at $Q^2>6$~GeV$^2$). This might
explain why observability of the $\Theta^+$-peak in the ZEUS data does
not need any additional restrictions for $W$ (or $x$) at higher $Q^2$.

The two quantities, $Q^2$ and $x$, are, generally, independent
variables, and a definite value of $x$ may (and, most probably,
does) select some states among various many-parton configurations
with the ``life time'' $\sim 1/\sqrt{Q^2}$, corresponding to the
used value of $Q^2$. If the $\Theta^+$ is not produced indeed at
too small $x$, it could mean, that typical states of the remnant
with presence of a quark having such $x$, may have properties
(\textit{e.g.}, quark energy distributions) that suppress the
$\Theta^+$-formation. Future experiments could study both $Q^2$
and $x$ dependencies of the $\Theta^+$-production in more detail,
to extract interesting information on structure of both the proton
remnant and the $\Theta^+$ itself. The related problems seem to be
worth of more detailed theoretical discussion elsewhere.

b) \textit{Some other positive data.}  Hadron remnants emerge in
all hard processes having hadron(s) in the initial state (DIS,
Drell-Yan pair production, production of high-$p_T$ hadrons, and
so on). But many-parton configurations may contribute also to
non-hard (soft) processes, \textit{e.g.}, through higher Fock
components of the initial hadrons. In framework of our hypothesis,
they also can generate the $\Theta^+$.

The corresponding manifestations of the $\Theta^+$-production
mechanism are possibly seen in preliminary data of the SVD
Collaboration. Its higher-statistics analysis~\cite{svd, svd-m} has
confirmed the earlier evidence for production of the $\Theta^+(1530)\,$
in nucleon-nucleon collisions~\cite{svd-e}. Moreover, preliminary data
of this collaboration provide distributions of the $\Theta^+$ in the
Feynman variable $x_F$ and in $p_T$.

The $x_F$-distribution for the $\Theta^+(1530)$ appears to be very
soft: no $\Theta^+(1530)$ is seen at $|x_F|>0.3$~\cite{svd-m,svd1},
while the observed yield of $\Lambda(1520)$ is much harder: it has
maximum at $|x_F|=0.4\,$~\cite{svd1}. Distribution of the $\Theta^+$
over $p_T$ is also soft (no $\Theta^+$ at $p_T> 1.5$~GeV)~\cite{svd1}.
The two distributions together imply that the total c.m.-momentum of
the produced $\Theta^+$ is not higher than 1.5~GeV, essentially smaller
than kinematically available (and than that for $\Lambda(1520)$). Such
properties look quite natural in framework of our hypothesis. If the
pentaquark $\Theta^+(1530)$ is indeed produced from many-parton
configurations, then we expect that accompanying hadron multiplicity
for the pentaquark $\Theta^+(1530)$ is higher than, say, for the
conventional hyperon $\Lambda(1520)$, which is mainly produced through
fewer-parton configurations. This makes the c.m.-energy, available
for the $\Theta^+(1530)$, to be smaller than that for the
$\Lambda(1520)\,$.

Similar evidence may come also from preliminary data of the HERMES
Collaboration~\cite{lor}. When the standard kinematic constraints,
used for the $\Theta^+$-observation, are appended by the
requirement of an additional detected pion, the signal/background
ratio essentially improves. It reaches 2:1, instead of 1:3 in the
published HERMES data~\cite{herm}. Up to now such a quite unexpected
result had no explanation, except trivial assumptions of a possible
artefact or an experimental error. Our hypothesis allows to understand
it in a reasonable way. The condition of an additional pion in the
detector enhances the role of events with higher multiplicity, where,
as we argue, the pentaquarks should be present with higher rate, in
respect to conventional hadrons, than in average events.

\subsection{Null data}

a) \textit{Low-energy measurements.} Two collaborations have
published impressive null results on the $\Theta^+(1530)\,$ at low
(or relatively low) initial energies.

One of them, the Belle Collaboration, used low-energy kaons,
produced in $e^+e^-$-annihi\-la\-tion, for secondary scattering
inside the detector~\cite{bel}. Physically, this experiment is similar
to that of \mbox{DIANA}~\cite{di1,bar}, so they may be directly
compared.  In both cases, one studies the charge exchange reaction
$K^+n\to pK_S$ inside a nucleus and looks for the $\Theta^+$  as an
intermediate resonance. Experimental results on this charge exchange
process can be directly interpreted in terms of the $\Theta^+$-width.
The Belle data do not show any $\Theta^+$-signal and provide the
restriction~\cite{bel}
\begin{equation} \label{bel-w} \Gamma_{\Theta^+} < 0.64~\rm{MeV}\,,
\end{equation} which may cast doubt on the value \begin{equation}
\label{w-ct} \Gamma_{\Theta^+}= (0.9\pm0.3)~\rm{MeV}\,, \end{equation}
extracted by Cahn and Trilling~\cite{ct} from earlier DIANA
analysis~\cite{di1}. However, that publication was incomplete,
and some additional assumptions on the background contributions
were necessary to extract the value (\ref{w-ct}). A more detailed
analysis of the higher-statistics DIANA data~\cite{bar} allows to find
the width with much less assumptions. Its new value~\cite{bar}
\begin{equation} \label{expG} \Gamma_{\Theta^+}=(0.36\pm0.11)~{\rm MeV}
\end{equation} looks extremely low \footnote{Of course, it is much
smaller than the experimental resolution. The method of extraction of
such small width is described in Ref.\cite{ct}.}, but does not
contradict to any other experimental result on $\Gamma_{\Theta^+}$. In
particular, with this new value, the current Belle data cannot exclude
existence of the $\Theta^+$.

Note that the value (\ref{expG}) is not yet quite understood
theoretically. Expectation of a narrow $\Gamma_{\Theta^+}$ appeared
when the $\chi$QSA provided evidence for calculational suppressions
in some coupling constants of exotic hadrons~\cite{dpp,ekp,pra}.
Moreover, traditional methods of $\chi$QSA, related to decomposition
in $1/N_c$, allow to find upper bounds for those coupling constants.
As a result, such methods predict $\Gamma_{\Theta^+} \leq15~{\rm
MeV}$~\cite{dpp}, which is very low in comparison with familiar widths
of baryonic resonances ($\sim50 - 100$~MeV). The ``model-independent
approach'' to the chiral quark soliton, using phenomenology of the
proton spin content and hyperon semileptonic decays, has recently
shown~\cite{mia} that the $\chi$QSA may be consistent with
$\Gamma_{\Theta^+} < 1$~MeV and even with the value~(\ref{expG}).
However, this approach cannot yet fix a particular theoretical value
for the width. Moreover, it has not explained a physical reason
for the suppression of $\Gamma_{\Theta^+}$.

A progress in understanding the $\Theta^+$-width seems to be related
with the Fock column picture of hadrons~\cite{dpF}. Starting from
the soliton of $\chi$QSA, the baryons were described at the light-cone
as a set of quarks (3 quarks appended by additional $q\overline{q}$
pairs) in the self-consistent mean chiral field. This way, with
various corrections, has lowered the theoretical value down to
$\Gamma_{\Theta^+}\approx 0.43$~MeV~\cite{cl,dd,cl1}, just comparable
with the experimental value (\ref{expG}). If this way is correct, it
opens possibility for physical explanation of such a tiny width. The
reason is participation of hadronic higher Fock components in the
process of the $\Theta^+$-decay. Evidently, it is similar to our
present hypothesis, which suggests participation of hadronic higher
Fock components in the process of the $\Theta^+$-production.

The other set of null data is given by the CLAS measurements of
photoproduction off proton and neutron (inside the deutron). Despite
impressive experimental statistics, treatment of several final states
has not revealed any reliable $\Theta^+$-signal~\cite{clas1,
clas2,clas3,clas4}. These results could pretend to disprove positive
evidences of LEPS.  However, the kinematical region of the CLAS
detector cannot completely overlap that of the other one (the
problem is mainly related with the forward direction), so a more
detailed comparison is necessary for definite conclusions.

Some current theoretical estimations~\cite{kwee,guz} predict even
lower cross sections for the $\Theta^+$-photoproduction, than the
experimental upper limits of CLAS. Of course, it is not clear at
present, whether such expectations may agree with the latest positive
observations of LEPS~\cite{leps}. However, today's predictions have
essential theoretical uncertainties, which touch, first of all, form
factors in exchange diagrams. The published calculations apply form
factors with properties familiar for conventional hadrons. However, the
form factors may be different, if the transition, say, $N\to\Theta^+$
goes mainly in 5-quark (or higher) configuration(s). According to ideas
of Feynman~\cite{feyn}, the more constituents has a system, the faster
should change (decrease) its form factor \footnote{The underlying
physics is rather transparent:  to change the motion of a whole system
of many constituents one should realize many interactions between all
those constituents, so to drag each of them.}. Then, according to our
hypothesis, photoproduction of the pentaquark $\Theta^+$ on the nucleon
should go as on a 5-quark system, with a form factor having steeper
$t$-dependence than in photoproduction of conventional hadrons. This
might produce the sharp angular distribution which would make mutually
consistent the observation of the $\Theta^+$, \textit{e.g.}, in the
forward-looking detector LEPS and its non-observation in the
less-forward-looking detector CLAS. In addition, if the CLAS data
really contain a small $\Theta^+$-signal, consistent with the published
bounds, this signal may be enhanced to the observable level due to
interference effects~\cite{adp}. Thus, the published CLAS data cannot
yet pretend to reject existence of the $\Theta^+$.

This is even more true for the COSY-TOF data. The new
measurements~\cite{tof} do not confirm earlier evidence of the same
collaboration for the $\Theta^+$-baryon in the reaction $pp\to pK^0
\Sigma^+$~\cite{tof0}. Instead, they set a strict limit for the
$\Theta^+$-production. However, their new upper boundary for the
cross section is still higher than the theoretical
estimation~\cite{pols} obtained before both new and older analyses.

b) \textit{High-energy measurements.} Null results have been published
also for several types of processes considered to be high-energy ones.
We begin with $e^+e^-\to\,$hadrons (and/or $\gamma\gamma\to\,$hadrons).
Such processes at high energies go mainly through the prompt production
of one quark-antiquark pair, which then hadronizes. This favorably leads
to purely mesonic final states (the additional $q\overline q$-pair may
be produced in a soft manner). Events with baryon-antibaryon pairs
need 3 prompt quark-antiquark pairs (\textit{i.e.}, two more pairs) and
provide only small part (phenomenologically less than, say, 1/10) of
all high-energy events. Production of pentaquark baryon(s) needs even
more, 5 prompt quark-antiquark pairs (two additional pairs in comparison
with conventional baryon events). Thus, one may expect that rate of
events with pentaquark(s) in high-energy $e^+e^-$ annihilation is less
than 1/10 of events with conventional baryons. Current experimental
data have not reached sufficient precision to observe them at such
level.

Similar (and even stronger) conclusions are also true for charmonium
decays, where modes with baryon-antibaryon pairs have the rate of order
1/100 in respect to purely meson modes. This implies that pentaquarks
in charmonium decays may be expected to have rate not more than 1/100
in regard to conventional baryons. Detailed analysis~\cite{as} shows
that the corresponding experimental data have strongly insufficient
precision.

In hadron-hadron collisions, there is the negative search for
the $\Theta^+$ by the SPHINX Collaboration~\cite{sph}. It is usually
claimed to reject the SVD positive result~\cite{svd,svd-m,svd1} (the two
collaborations collected their data at the same Serpukhov proton beam of
70 GeV). However, the SPHINX detector is sensitive mainly to the area
of diffraction dissociation (excitation) of the initial proton. At the
Serpukhov energy, this process can be well described by an effective
pomeron interacting with constituent quarks, without changing the number
of constituents. Therefore, our hypothesis implies strong suppression of
multiquark baryons in the SPHINX experiment. The suppression may be
weaker in the SVD case, where contribution of many-quark fluctuations
could provide a more noticeable effect, than in a special case of
diffraction dissociation. The SVD and SPHINX measurements have been
compared also from a different, ``instrumental'' viewpoint by the
authors of Ref.\cite{ChL}. They also conclude that the SVD data are
more favorable to search for the narrow baryon $\Theta^+(1530)\,$.

Large sets of high-energy data for the $\Theta^+$-search in
hadron-hadron collisions has been obtained at hadronic colliders.
Their detectors usually tag events with high values of the
transverse energy. Such events are mainly initiated by partons
knocked-out from initial hadrons, which then hadronize essentially
as few-parton systems. Multiquark hadrons should be very rare in
their hadronization products. For the same events, our hypothesis
suggests to expect a higher rate of exotic hadron production in
remnants of the initial hadrons. However, these remnants are
practically lost for the FNAL detectors CDF and/or D0. The same is
true for the main RHIC detectors, at least, in symmetric collisions
$p+p\,,~Cu+Cu\,$ and $\,Au+Au\,$. Up to now, dedicated investigation
of the hadron remnants has not been planned for those collider
detectors in any approved experiments.

Thus, the current high-energy data, which do not see the
$\Theta^+$, cannot be decisive for the pentaquark problem, even
despite their greater statistics. In framework of our hypothesis,
the reason is that they investigate kinematical regions where
multiquark hadrons are produced with very low rate.

\subsection{Suggestions for further experiments}

Though we do not review all existing data, the above examples show that
our hypothesis about exotics production mechanism allows to reconcile
current positive and null experiments on the $\Theta^+$-baryon. Even
more, it suggests new approaches for studying multiquark hadrons. The
corresponding experiments should be oriented to hadrons produced from
many-quark systems. As we explained, a clear example of such systems is
provided by hadron remnants in hard processes with initial hadrons.
At current symmetric configurations of the colliding beams, the
remnants escape into tubes of the storage rings. Thus, to investigate
them, one would need to essentially modify the existing detectors,
to register forward-going hadrons with very high rapidity.

Directions of necessary modification may be demonstrated by the
BRAHMS forward spectrometer. It has recently presented the first
measurements of high-rapidity production of pions, kaons, protons, and
their antiparticles in $pp$-collisions at RHIC ($y_{\,had}\approx3$,
while $y_{\,beam}=5.4$ at $\sqrt{s}=200$~GeV)~\cite{brahms}. The data
for pions and kaons agree with calculations up to Next-To-Leading-Order
in the perturbative QCD, but (anti)protons in that region show yet
unexplained deviations from expected distributions (see
Ref.~\cite{brahms} for more detailed comparison of data \textit{vs.}
calculations). Baryon spectroscopy in this region may also reveal
unexpected features.

Another approach could be applied, if one used, say, at the FNAL
collider, an asymmetric configuration with $p$ and $\overline p$
beams of unequal energies.  Then, the Lorentz boost would increase
the angular dimension for the remnant of the less-energetic hadron,
so it could be seen even in the existing detectors, without their
modifications. Such a way would allow to investigate kinematical
regions in high-energy hadronic collisions, which have not been seen
up to now and which could provide, as we expect, higher rate of
exotics production, than in the published collider experiments.

Note, that there exist asymmetric colliders. For instance, the
$B$-factories at SLAC and KEK are the asymmetric $e^+e^-$-colliders.
But they study the process $e^+e^-\to\,\,$hadrons, where, as we
explained, exotic baryon production should be an exotic process
indeed. The HERA facility at DESY is also just an asymmetric
collider. However, its asymmetry (the proton momentum much higher
than the electron one) is such that the detectors ZEUS and H1 favor
investigations of the current fragmentation region, while the target
fragmentation region, directly related to the proton remnant, is
strongly shrunk. We expect that the HERA run with the less energetic
proton beam could allow the ZEUS detector (and may be H1 as well) to
study in more detail the proton remnant contributions.  A very
interesting and perspective direction of studies could be
investigation of $x$- and $Q^2$-dependencies of the
$\Theta^+$-production in DIS. In particular, we expect that, with
$Q^2$ growing, the ratio of exotic/conventional hadron yields should
logarithmically increase, in analogy with logarithnic increase of
the small-$x$ hadronic structure functions.

Many-quark systems may emerge also in soft processes, due to short-term
fluctuations of initial hadron(s). However, their relative contribution
is usually rather small and, thus, should provide rather small cross
section for the exotics production, as compared to conventional hadron
production \footnote{Note, that hard processes, directly related to the
short-term fluctuations, also have small cross sections, in comparison
with soft processes. Separation of a hard process from much more
copious soft background requires some special selection conditions,
\textit{e.g.}, high-$p_T$ or high-$Q^2$ events.}. To increase an
exotics signal/background ratio one needs to apply some selections
enhancing the role of many-parton fluctuations.  A possible way could
be to study events with high (total or charged) hadron multiplicity.
Of course, such events have larger combinatorial background. But we
expect that the relative yield of the exotic $\Theta^+$-baryon
\textit{vs.} a conventional baryon, say, $\Lambda(1520)\,$, should
be enhanced for such events in comparison to the total set of events.
Properties of the $\Theta^+$-production, as observed by the SVD
Collaboration~\cite{svd,svd-m,svd-e,svd1}, show that also useful for
the exotics observation might be restriction to relatively low energy
of the $(KN)$-pair (as we have explained, this is also related to the
enlarged multiplicity).

The two latter points may be essential in evaluating null-result
experiments for exotics searches. Some of those experiments, because
of ``technical'' reasons, apply restrictions on particle energies
and/or on multiplicity of selected events. The above discussion
demonstrates that these restrictions can suppress (or enhance!) the
relative exotics production. Such possibilities should be taken into
account in planning future experiments.

\section{Conclusions and discussions}

Let us summarize the above considerations. On the base of some
positive experimental data, we assume that exotic (and, more generally,
multiquark) hadrons are mainly generated from many-quark partonic
configurations, which may emerge either as short-term fluctuations of
initial hadrons in any hadronic process, or as hadron remnants in hard
processes (which are usually considered to be just remnants of the
short-term fluctuations).  Usually, manifestations of short-term
fluctuations (or, of higher Fock components) of the initial hadron(s)
are related with such hard hadronic processes as DIS, Drell-Yan pair
and/or high-$p_T$ hadron production, and some others. Our hypothesis
is a generalization of the suggestion of Karliner and
Lipkin~\cite{kl,lip}, who assumed the $\Theta^+$-production mainly
going through a special multiquark resonance. In difference, we do
not stick to particular resonance(s) and admit contributions from
any many-quark state.

If our hypothesis proves true, the multiquark production in processes
with initial hadrons presents a new kind of hard processes, in
addition to such familiar hard processes as DIS, high-$p_T$ hadron
production, Drell-Yan pair production, and so on. At first sight, they
are essentially different: exotics production has not any continuous
``regulator'' of the hardness, which exists for many other processes
(values of $Q^2\,$, $p_T\,$, Drell-Yan mass, and so on). However, in
production of heavy quarkonium or heavy quark hadrons the hardness
``regulator'' is not continious; it is fixed by the heavy quark mass.
We expect that the  characteristic time scale for production of
multiquark hadrons should also be shorter than for soft production of
the conventional hadrons. If this process is related indeed with higher
Fock components, the exotics time scale should be somehow fixed by the
minimal number of quarks in the exotic hadron.

We have demonstrated above that our assumption is really able to
explain, at least qualitatively, why the $\Theta^+$ has been seen
in some experiments and not in others. Moreover, it allows to overcome
seeming inconsistencies between experiments, which are similar at first
sight, and to understand such peculiar features of positive experiments
as, \textit{e.g.}, essential change of signal/background ratio with
relatively small change of registration conditions.

We can further suggest new experiments that might be decisive to check
(and hopefully confirm) existence of the $\Theta^+$ and other multiquark
hadrons. Most important for this purpose could be studies of the hadron
remnants in hard processes. Up to now, structure and evolution of those
remnants have not been experimentally investigated and/or theoretically
understood.

If our assumptions appear correct, experiments with production and
investigation of multiquark hadrons could provide new, very interesting
and important information about structure of conventional hadrons and
about properties of their short-term fluctuations. In particular, it
might help to understand structure and properties of higher Fock
components for both conventional and multiquark hadrons.

Studies of multiquark hadrons will reveal, of course, a new hadronic
spectroscopy, unobserved till now. They may also give a fresh look at
constituent quarks. As is known, the constituent quarks are absent in
the QCD Lagrangian, they emerge only as efficient objects (possibly,
similar to quasiparticles in solid-state physics). Properties of such
objects may depend on the environment. Therefore, the effective masses
and couplings of the constituent quarks might be different in the
conventional hadrons and in multiquark ones. If this phenomenon were
discovered, it would strongly advance understanding the nature of the
constituent quarks~\footnote{Relation between exotic hadrons and the
constituent quark model has been recently discussed from another
viewpoint by Lipkin~\cite{lip1}.}. All the above prospects are only
examples of the progress that could be related with exotic hadrons.

\section*{Acknowledgments}

We thank A. Airapetian, S, Chekanov, D. Diakonov, A. Dolgolenko,
M. Karliner, A. Kubarovsky, H. Lipkin, V. Petrov, M. V. Polyakov, and
R. Workman for useful discussions on various sides of the exotics
problem. Ya.A. thanks the Ruhr-Universit\"at-Bochum for hospitality
at some stages of this work.  The work was partly supported by the
Russian State Grant RSGSS-1124.2003.2, by the Russian-German
Collaboration Treaty (RFBR, DFG), by the COSY-Project J\"ulich, by
Verbundforschung ``Hadronen und Kerne'' of the BMBF, by Transregio/SFB
Bonn, Bochum, Giessen of the DFG, by the U.~S.~Department of Energy
Grant DE--FG02--99ER41110, by the Jefferson Laboratory, and by the
Southeastern Universities Research Association under DOE Contract
DE--AC05--84ER40150.

\end{document}